\documentclass[aps,prb,twocolumn,showpacs,groupedaddress,preprintnumbers]{revtex4-1}

\usepackage{graphicx}

\begin{document}

\preprint{\textit{Published version of this paper can be found at} Phys. Rev. B \textbf{83}, 054509 (2011).}

\title{Determination of the superconducting gap in near optimally doped Bi$_2$Sr$_{2-x}$La$_{x}$CuO$_{6+\delta}$ ($x\sim0.4$) from low-temperature specific heat}



\author{Yue Wang$^{1}$}
\email[]{yue.wang@pku.edu.cn}

\author{Zhi-Yong Liu,$^2$ C. T. Lin$^3$}

\author{Hai-Hu Wen$^{4,5}$}
\email[]{hhwen@nju.edu.cn}

\affiliation{$^1$State Key Laboratory for Mesoscopic Physics and School of Physics, Peking University, Beijing 100871, People's Republic of China}

\affiliation{$^2$Department of Physics, Shanghai University, Shanghai 200444, People's Republic of China}

\affiliation{$^3$Max-Planck-Institut f\"{u}r Festk\"{o}rperforschung, Heisenbergstrasse 1, D-70569 Stuttgart, Germany}

\affiliation{$^4$National Laboratory for Superconductivity, Institute of Physics and Beijing National Laboratory for Condensed Matter Physics, Chinese Academy of Sciences, Beijing 100190, People's Republic of China}

\affiliation{$^5$National Laboratory of Solid State Microstructures and Department of Physics, Nanjing University, Nanjing 210093, People's Republic of China}


\date{\today}

\begin{abstract}
Low-temperature specific heat of the monolayer high-$T_c$ superconductor Bi$_2$Sr$_{2-x}$La$_{x}$CuO$_{6+\delta}$ has been measured close to the optimal doping point ($x\sim0.4$) in different magnetic fields. The identification of both a $T^2$ term in zero field and a $\sqrt{H}$ dependence of the specific heat in fields is shown to follow the theoretical prediction for $d$-wave pairing, which enables us to extract the slope of the superconducting gap in the vicinity of the nodes ($v_{\Delta}$, which is proportional to the superconducting gap $\Delta_0$ at the antinodes according to the standard $d_{x^2-y^2}$ gap function). The $v_{\Delta}$ or $\Delta_0$ ($\sim12$~meV) determined from this bulk measurement shows close agreement with that obtained from spectroscopy or tunneling measurements, which confirms the simple $d$-wave form of the superconducting gap.
\end{abstract}


\maketitle



\section{INTRODUCTION}

In the study of high-$T_c$ cuprate superconductors, low-temperature specific heat (LTSH) has proven to be a helpful tool in identifying the $d$-wave symmetry of the superconducting gap.~\cite{Hussey02,Fisher07} In zero magnetic field, the quasiparticle density of states of a $d$-wave superconductor shows a linear energy dependence owing to the presence of line nodes, which is expected to give rise to a $T^2$ temperature-dependent electronic specific heat. In magnetic field $H$, an energy shift (Doppler shift) to the nodal quasiparticle spectrum becomes important, and the electronic specific heat of a $d$-wave superconductor is predicted to have a characteristic $\sqrt{H}T$ dependence.~\cite{Volovik93} In LTSH experiments on high-$T_c$ cuprates, although it still seems to be somewhat controversial in the identification of the $T^2$ electronic specific heat in zero field (partly owing to its small magnitude),~\cite{Hussey02,Fisher07,Loram01} the $\sqrt{H}T$-dependent specific heat in fields has been widely recognized in both YBa$_2$Cu$_3$O$_{7-\delta}$ (YBCO) and La$_{2-x}$Sr$_x$CuO$_4$ (LSCO),~\cite{Hussey02,Fisher07} hence providing the bulk evidence for $d$-wave pairing in these prototypical high-$T_c$ compounds.

Recently, more detailed work showed that a key parameter of the $d$-wave superconducting gap, i.e., the gap slope in the vicinity of the nodes $v_{\Delta}$, also can be quantitatively determined from LTSH.~\cite{Wang01,Wen05,Wang07} This makes LTSH an effective measure of the superconducting gap maximum at the antinodes, $\Delta_0$, which is proportional to the $v_{\Delta}$ according to the standard $d_{x^2-y^2}$ gap form $\Delta=\Delta_0\cos(2\phi)$, as exemplified by experiments on LSCO across a wide doping range.~\cite{Wang07} The determination of the superconducting gap from LTSH has the merits that it is sensitive to the gap structure near the nodes by probing the low-energy quasiparticle excitations and it reflects the bulk property of the sample. A comparison of it with more traditional gap measurements such as spectroscopy or tunneling may give valuable insights into the superconducting state.

In light of the above, we present in this paper the LTSH experiment on a monolayer high-$T_c$ cuprate superconductor Bi$_2$Sr$_{2-x}$La$_{x}$CuO$_{6+\delta}$ (La-Bi2201).Up to now, the field-dependent LTSH study to identify the $d$-wave pairing effect has seemed to be lacking for this material.~\cite{Hussey02,Fisher07} On a single-crystal sample with a doping level close to the optimal ($x\sim0.4$), we have resolved a $T^2$ electronic specific heat in zero field and a $\sqrt{H}T$-dependent electronic specific heat in fields, both in conformity with the $d$-wave pairing theory. This has then allowed us to quantitatively determine the $v_{\Delta}$ and a superconducting gap amplitude $\Delta_0$ of $\sim$12~meV for the sample. The $v_{\Delta}$ or $\Delta_0$ is found to show reasonable agreement with that determined by angle-resolved photoemission spectroscopy (ARPES) or scanning tunneling microscopy (STM) for La-Bi2201 at the optimal doping level.~\cite{Harris97,Kondo07,Wei08,Ma08,Meng09}

\section{EXPERIMENT}

The La-Bi2201 single crystal with $x\sim0.4$ and a mass of 10.2~mg was grown by the traveling solvent floating-zone method as reported elsewhere.~\cite{Liang04} The $T_c$ of the sample is 28~K, defined by the onset of the magnetization in an ac susceptibility measurement.~\cite{Note1} The LTSH measurement was performed using a thermal relaxation method, as described in detail previously.~\cite{Wen04} It was done in an Oxford Maglab cryogenic system in the temperature range $2.5-10$~K and in magnetic fields up to 12~T ($H\parallel~c$ axis). The heat capacity of the addenda was measured separately and was $\sim$30$\%$ of that of the sample in the measured temperature range. The chip thermometer has been calibrated in fields and the heat capacity of the addenda is mostly due to phonons and hence shows nearly no field dependence.~\cite{Wen04}

\section{DATA AND ANALYSIS}

\begin{figure}
\includegraphics[scale=0.35]{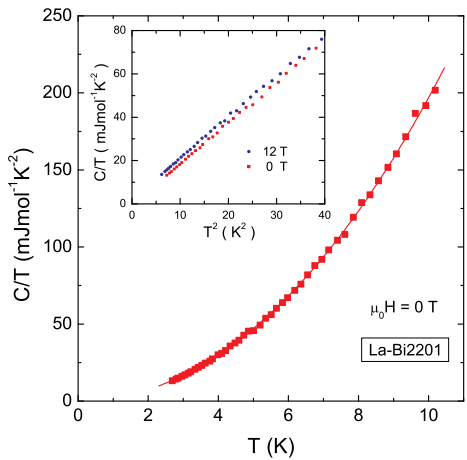}
\caption{\label{figure1}(Color online) Specific heat of the sample in zero magnetic field, plotted as $C/T$ vs $T$ (squares). The solid line is the fit to $C=\beta_0T^3+\eta_0T^5$. The inset shows the specific heat below 6.5~K in zero field and in 12~T, plotted as $C/T$ vs $T^2$.}
\end{figure}

Figure~\ref{figure1} shows the specific heat of the sample in zero field. We found it could be well fitted by the formula $C=\beta_0T^3+\eta_0T^5$ with $\beta_0$ and $\eta_0$ representing the coefficient of the phonon $T^3$ term and $T^5$ term respectively. The fit is shown by a solid line in Fig.~\ref{figure1} and Table~\ref{table1} lists the parameters $\beta_0$ and $\eta_0$. This indicates that in zero field the specific heat comes mainly from phonon excitations. Note that the $\beta_0$ for the present sample is roughly five or eight times larger than that reported for optimally doped YBCO or LSCO, respectively,~\cite{Moler97,Fisher00} reflecting a substantially greater phonon specific heat for La-Bi2201. From $\beta_0$ the Debye temperature $\Theta_D$ is calculated as 226~K. A residual linear specific heat, $\gamma(0)T$, is usually observed for YBCO and LSCO in zero field,~\cite{Hussey02,Fisher07,Wang07} while for our La-Bi2201 crystal, the above fit suggests that such a term should be close to zero. This can be further illustrated by the inset of Fig.~\ref{figure1}. In plot of $C/T$ vs $T^2$, it is seen that the 0-T data show a nearly zero intercept on the $C/T$ axis at $T=0$~K, indicating a negligible $\gamma(0)T$ term. It is worth noting that a nearly zero $\gamma(0)$ has also been reported previously for La-free Bi2201 and Bi$_2$Sr$_2$CaCu$_2$O$_{8+\delta}$ (Bi2212) single crystals.~\cite{Chakraborty89,Urbach89} The reason why the $\gamma(0)T$ is shown for YBCO and LSCO while not for most Bi-based cuprates at the optimal doping level is presently not clear.~\cite{Fisher07} In the inset of Fig.~\ref{figure1}, we have also shown the specific heat in a field of 12~T. A nearly parallel shift to the zero-field data, though small, can be readily seen. This gives a finite intercept on the $C/T$ axis for 12-T data and hence implies that there is a field-induced linear-$T$ specific heat, $\gamma(H)T$.

\begin{figure}
\includegraphics[scale=1]{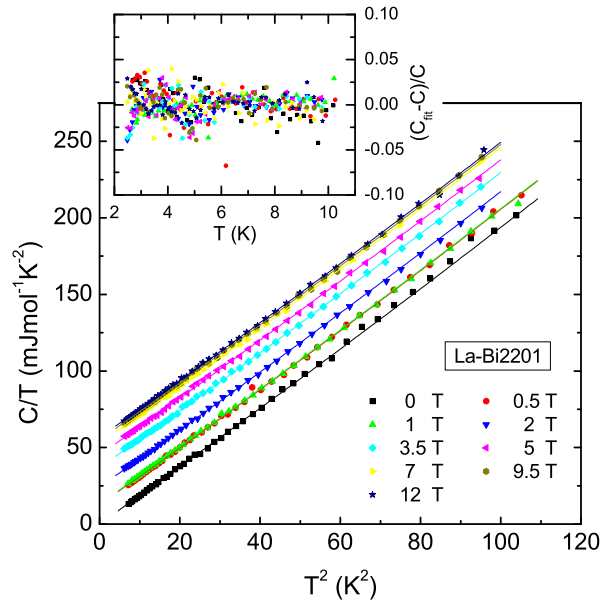}
\caption{\label{figure2}(Color online) Global fit (solid lines) to the specific heat in fields up to 12~T (symbols), plotted as $C/T$ vs $T^2$. All data are fit simultaneously to Eqs.~(\ref{eq:1}) and (\ref{eq:2}) with $\beta$ and $\eta$ kept the same for all $H$. For clarity, the interval between data sets at different $H$ has been enlarged and the in-field data and fit are shown as $C/T=20\gamma(H)+\beta T^2+\eta T^4$. The inset shows the difference between the data and the fit, $(C_{\mathrm{fit}}-C)/C$ vs $T$.}
\end{figure}

One may note that no $\alpha$$T^2$ term expected for a $d$-wave superconductor is contained in the above fitting to the zero-field data. This is because we found that if we allowed it to be in the fit, we would obtain a negative value of the coefficient $\alpha$. In fact, previous studies have shown that it is usually difficult to identify the $T^2$ term by fitting the zero-field data alone because this term is expected to be rather small and could be easily concealed by the large phonon specific heat.~\cite{Moler97,Fisher00} Using the data in different fields together to more accurately determine the phonon contribution is required to give a reliable evaluation of the $\alpha$$T^2$ term.~\cite{Fisher07} Therefore, to correctly identify the $T^2$ term in zero field and determine the linear-$T$ specific heat in fields, we have performed a global fit to all the LTSH data,~\cite{Moler97} in which the zero-field data set is fit to
\begin{equation}
C(T,0)=\alpha T^2+\beta T^3+\eta T^5\label{eq:1}
\end{equation}
and the in-field data sets ($\mu_0H\geq0.5$~T) are fit to
\begin{equation}
C(T,H)=\gamma(H) T+\beta T^3+\eta T^5.\label{eq:2}
\end{equation}
In this global fit, all data sets are fit simultaneously to Eqs.~(\ref{eq:1}) and (\ref{eq:2}) with a weighted-least squares criterion [minimizing $\sum(\frac{C_\mathrm{fit}-C_i}{C_i})^2$].~\cite{Note2} The phonon coefficients $\beta$ and $\eta$ are kept as the same for all data while no assumptions on the magnitude of $\alpha$ or on the field-dependent form of $\gamma(H)$ are made. Figure~\ref{figure2} presents the global-fit result, where, for clarity, the interval between data sets at different $H$ has been enlarged 20 times according to the fit. One can see the data are reproduced well by solid fit lines, showing the reasonable quality of the fit.  The inset of Fig.~\ref{figure2} shows the difference between the data and the fit, from which the total rms deviation of the data from the fit was calculated to be 1.1$\%$.

\begin{table}
\caption{\label{table1}Fit parameters in the present LTSH study. $\beta_0$ ($\beta$), $\eta_0$ ($\eta$), $\alpha$, and $\gamma(H)$ are in units of mJ~mol$^{-1}$~K$^{-n}$ with $n=4$, 6, 3, and 2, respectively.}
\begin{ruledtabular}
\begin{tabular}{ccccccc}
Fit               & Parameters                                           \\
\hline
$H=0$             & $\beta_0=1.86\pm0.01$, $\eta_0=1.14\times10^{-3}$      \\
\hline
Global fit        & $\beta=1.84\pm0.01$, $\eta=8.93\times10^{-4}$,         \\
                  & $\alpha=0.10\pm0.04$, $\gamma(H)=(0.90\pm0.06)\sqrt{H}$          \\
\end{tabular}
\end{ruledtabular}
\end{table}

\begin{figure}
\includegraphics[scale=0.35]{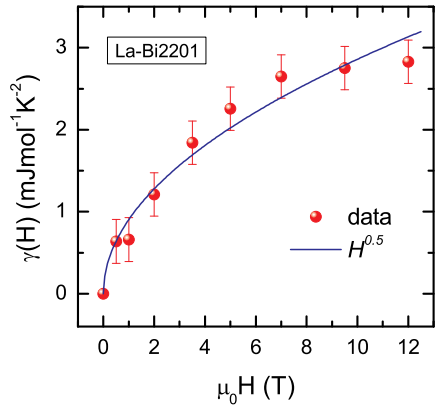}
\caption{\label{figure3}(Color online) Coefficient of the field-induced linear-$T$ specific heat, $\gamma(H)$, obtained from the global fit (circles). Error bars are at a statistical 90$\%$ confidence level. The line is the fit to $\gamma(H)=A\sqrt{H}$ ($A=0.90$ mJ~mol$^{-1}$~K$^{-2}$~T$^{-0.5}$) for $d$-wave pairing.}
\end{figure}

The parameters yielded from this fit are listed in Table~\ref{table1}, with 90$\%$ statistical confidence levels. Unlike the single fit to zero-field data, the $\alpha$ determined from the global fit is positive, supporting the existence of a $T^2$ term in zero field. It is noted that the size of the $\alpha$ is comparable to that determined for optimally doped YBCO or LSCO.~\cite{Moler97,Fisher00} Figure~\ref{figure3} plots the $\gamma(H)$ at each field. Plotted together is a solid line showing the fit to $\gamma(H)=A\sqrt{H}$ based on the theoretical prediction for a $d$-wave superconductor.~\cite{Volovik93} It is seen that the solid line gives a reasonable description to the $\gamma(H)$, suggesting that the field-induced specific heat essentially follows the predicted $\sqrt{H}T$ dependence, in line with the observation in YBCO and LSCO.~\cite{Hussey02,Fisher07,Wang07} This, alongside the presence of an $\alpha$$T^2$ term in zero field, provides evidence for the dominance of $d$-wave symmetry in the superconducting gap in optimally doped La-Bi2201. Note that although the $d$-wave pairing has been probed for La-Bi2201 by other experimental techniques such as ARPES,~\cite{Harris97,Kondo07,Wei08,Ma08,Meng09} present LTSH findings confirm that it is a bulk property of the sample.

A more quantitative analysis can be made. Within a $d$-wave framework, the magnitude of the $\alpha$$T^2$ term depends on the density of states near the gap nodes, and hence correlates with the nodal gap slope, $v_\Delta$, which characterizes how steeply the gap opens around the nodes. Specifically, the coefficient $\alpha$ can be expressed as
\begin{equation}
\alpha=\frac{18\zeta(3)}{\pi}\frac{k^3_B}{\hbar^2}\frac{nV_{\mathrm{mol}}}{d}\frac{1}{v_Fv_\Delta},\label{eq:3}
\end{equation}
where $\zeta(3)\simeq1.2$, $k_B$ is Boltzmann's constant, $\hbar$ is Planck's constant, $n$ is the number of CuO$_2$ planes per unit cell, $d$ is the unit-cell size along the $c$ axis, $V_\mathrm{mol}$ is the mole volume of the unit cell, and $v_F$ is the Fermi velocity at the nodes.\cite{Chiao00,Vekhter01,Hussey02} Similarly, as the field-induced specific heat $A\sqrt{H}T$ of a $d$-wave superconductor comes from the Doppler shift of the quasiparticle states (owing to the superfluid flow around the vortices) in the vicinity of the nodes,~\cite{Volovik93} its magnitude is also sensitive to the $v_\Delta$. The relation between the prefactor $A$ and the $v_\Delta$ is given by
\begin{equation}
A=\frac{4k_{B}^{2}}{3\hbar}\sqrt{\frac{\pi}{\Phi_{0}}}\frac{nV_{\mathrm{mol}}}{d}\frac{a}{v_{\Delta}},\label{eq:4}
\end{equation}
where  $\Phi_0$ is magnetic flux quantum and $a=0.465$ for a triangular vortex lattice.\cite{Chiao00,Vekhter01,Hussey02} Equations~(\ref{eq:3}) and (\ref{eq:4}) explicitly show that with the $\alpha$ or $A$ determined in the experiment, one can in principle extract the $v_\Delta$ from LTSH with no adjustable parameters, which is quite helpful because, according to the standard $d$-wave form $\Delta=\Delta_0\cos(2\phi)$, one can further obtain a more familiar quantity, i.e., the superconducting gap maximum $\Delta_0$ at the antinodes, via $2\Delta_0=\hbar k_Fv_\Delta$, where $k_F$ is the Fermi wave vector along the nodal direction $(0,0)-(\pi,\pi)$.

For La-Bi2201, $n=2$, $d=24.4$~\AA, and $V_\mathrm{mol}=106.8$~cm$^3$. A recent ARPES experiment on optimally doped La-Bi2201 showed that along the nodal direction $v_F\simeq2.7\times10^{5}$~m~s$^{-1}$, and $k_F\simeq0.74$~\AA$^{-1}$.~\cite{Hashimoto08} Using these parameters and the $\alpha$ or $A$ as shown in Table~\ref{table1}, we obtain $\Delta_0\simeq13\pm5$~meV or $\simeq10.4\pm0.7$~meV according to Eq.~(\ref{eq:3}) or (\ref{eq:4}), respectively. It is seen that both determinations of the $\Delta_0$, which are largely independent, show a fairly good consistency, although the former has a larger uncertainty owing to the larger uncertainty in determining the $\alpha$. This further validates the analysis of the LTSH data within the $d$-wave pairing theory. From $\Delta_0\simeq10.4\pm0.7$~meV, which is determined from the field dependence of the specific heat and hence has a higher degree of accuracy, one can have a gap-to-$T_c$ ratio $\Delta_0/k_BT_c\simeq4.3\pm0.3$ for our sample, which is larger than the weak-coupling $d$-wave BCS prediction~\cite{Won94} $\Delta_0/k_BT_c=2.14$ and comparable to the values found in other optimally doped high-$T_c$ cuprates such as YBCO ($\sim4.3$ in Ref.~\onlinecite{Nakayama07}), LSCO ($\sim4.9$ in Ref.~\onlinecite{Wang07}), and Bi2212 ($\sim4.5$ in Ref.~\onlinecite{Ding96}).

\section{DISCUSSION}

\begin{figure}
\includegraphics[scale=0.4]{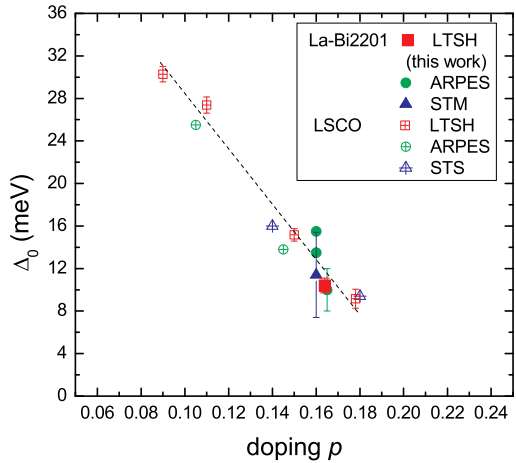}
\caption{\label{figure4}(Color online) Comparison of the superconducting gap $\Delta_0$ in near optimally doped La-Bi2201 obtained from the present LTSH study [according to Eq.~(\ref{eq:4}), solid square] and from some ARPES (solid circles) (Refs.~\onlinecite{Harris97,Wei08,Ma08,Meng09}) or STM (solid triangle) (Ref.~\onlinecite{Ma08}) studies. The $\Delta_0$ from LTSH is inferred from the nodal gap slope $v_{\Delta}$ under the standard $d$-wave form $\Delta=\Delta_0\cos(2\phi)$. The $\Delta_0$ from ARPES was measured at the antinodes in Ref.~\onlinecite{Harris97} or determined with the observation of a standard $d$-wave superconducting gap in Refs.~\onlinecite{Wei08,Ma08,Meng09}. To gain a better perspective, the comparison of the $\Delta_0$ in LSCO obtained from LTSH (crossed squares) (Refs.~\onlinecite{Wen05,Wang07}), ARPES (crossed circles) (Ref.~\onlinecite{Shi08}), and STS (crossed triangles) (Ref.~\onlinecite{Nakano98}) is also plotted in the doping range from slight overdoping to underdoping. The dashed line is a guide to the eye. See the text for details.}
\end{figure}

Focusing on La-Bi2201 at the optimal doping level ($x\sim0.4$), it is instructive to compare the $\Delta_0$ determined here to what has been measured by other experimental techniques such as ARPES and STM. In an early ARPES experiment,~\cite{Harris97} an antinodal superconducting gap of $10\pm2$~meV was observed, which shows good agreement with the LTSH result. This agreement confirms that the superconducting gap essentially follows the simple $d$-wave form because in ARPES the $\Delta_0$ is measured directly at the antinodes while in LTSH it is derived from the nodal gap slope $v_\Delta$ with the assumption $\Delta=\Delta_0\cos(2\phi)$. In several recent ARPES studies, it was found that from nodes to antinodes the superconducting gap can indeed be well fitted by the simple $d$-wave form with $\Delta_0$ of 13.5~meV (Refs.~\onlinecite{Wei08,Ma08}) or 15.5~meV.~\cite{Meng09} In a recent STM experiment,~\cite{Ma08} an averaged superconducting gap of $11.4\pm4$~meV was determined from spatial conductance maps. We can see these reports on $\Delta_0$ agree quantitatively with the LTSH result as well. In Fig.~\ref{figure4} we have plotted the $\Delta_0$ for near optimal-doped La-Bi2201 inferred from LTSH [according to Eq.~(\ref{eq:4})] and aforementioned ARPES and STM studies,~\cite{Harris97,Wei08,Ma08,Meng09} where the doping level of the sample ($p\sim0.16$) is determined by the empirical relation~\cite{Ando00,Luo08} between $p$ and the reported La content.

Unlike the above-compared ARPES experiments,~\cite{Harris97,Wei08,Ma08,Meng09} Kondo \textit{et al.} reported an antinodal gap as large as $\sim$40~meV below $T_c$ in optimally doped (Pb,La)-Bi2201 and ascribed it to the presence of a pseudogap with a nonsuperconducting origin.~\cite{Kondo07} To compare LTSH with such an ARPES experiment where a different energy scale other than the superconducting gap seemed present at the antinodes,~\cite{Millis06} one needs to focus directly on the $v_\Delta$ near the nodes. In fact, in the work of Kondo \textit{et al.}, it was shown that an extrapolation of the gap measured around the nodes under the standard $d$-wave form would yield a gap amplitude of $\sim$15~meV at the antinodes,~\cite{Kondo07} suggesting that the $v_\Delta$ they probed is also comparable to that obtained in LTSH. This further helps reveal that, whether a large pseudogap was detected or not at the antinodes below $T_c$, in recent ARPES measurements~\cite{Kondo07,Wei08,Ma08,Meng09} the superconducting gap determined near to the nodes exhibited consistent behavior, i.e., a quite similar nodal gap slope.

The above comparison shows that LTSH provides a reliable determination of the superconducting gap in La-Bi2201 at the optimal doping level, mimicking our previous observation in LSCO at various doping levels. As pointed out earlier, in recent work we have also determined the superconducting gap in LSCO across a wide doping range by probing the field dependence of the LTSH.~\cite{Wen05,Wang07} In the overdoped regime, the obtained $\Delta_0$ was found to show a quantitative agreement with that derived from other experimental techniques such as scanning tunneling spectroscopy (STS).~\cite{Wang07} In the doping range from slight overdoping to underdoping, Fig.~\ref{figure4} shows that in LSCO the $\Delta_0$ obtained from LTSH is also well confirmed by STS or recent ARPES measurements.~\cite{Nakano98,Shi08} Hence this illustrates the applicability of extracting the superconducting gap from LTSH across nearly the whole superconducting phase diagram of LSCO. Note that it also indicates a growing of the $v_\Delta$ below optimal doping toward the underdoped regime,~\cite{Wen05} and that the superconducting gap basically retains the simple $d$-wave form with diminishing $p$ (at least down to $p\sim0.11$), which is of importance particularly in the context of a recent debate on the gap property of the underdoped high-$T_c$ cuprates.~\cite{Millis06} In view of these findings in LSCO, we believe it would be appealing in the future to explore the LTSH of La-Bi2201 at more doping levels, with the expectation of acquiring the doping dependence of the superconducting gap in this archetypal cuprate from a bulk, thermodynamic method.

\section{CONCLUSION}

In summary, we have measured the LTSH in different magnetic fields in near optimally doped La-Bi2201. By performing a global analysis of the data, we have identified a $T^2$ electronic specific heat for $H=0$ and a $\sqrt{H}T$ electronic specific heat for $H\neq0$, as suggested for a $d$-wave superconductor. This provides the bulk evidence for $d$-wave pairing in optimally doped La-Bi2201, joining previous findings in YBCO and LSCO.~\cite{Hussey02,Fisher07,Wang07} Moreover, we have quantitatively determined the slope of the superconducting gap $v_\Delta$ near the nodes and its amplitude $\Delta_0$ ($\sim12$~meV) at the antinodes according to the standard $d_{x^2-y^2}$ form, which show close agreement with the results of ARPES and STM.~\cite{Harris97,Kondo07,Wei08,Ma08,Meng09} As shown in LSCO,~\cite{Wen05,Wang07} the present study further demonstrates the virtue of LTSH as a bulk method to probe the superconducting gap near the nodes in high-$T_c$ cuprates.

\begin{acknowledgments}

Technical assistance from Jing Wang is gratefully acknowledged. This work is supported by NSFC, CAS (Project ITSNEM), and MOST of China (973 Program, Nos. 2006CB601000 and 2011CBA00100).

\end{acknowledgments}

\bibliography{Bi2201LTSHRef}

\begin{thebibliography}{10}%
\makeatletter
\providecommand \@ifxundefined [1]{%
 \ifx #1\undefined \expandafter \@firstoftwo
 \else \expandafter \@secondoftwo
\fi
}%
\providecommand \@ifnum [1]{%
 \ifnum #1\expandafter \@firstoftwo
 \else \expandafter \@secondoftwo
\fi
}%
\providecommand \enquote [1]{``#1''}%
\providecommand \bibnamefont  [1]{#1}%
\providecommand \bibfnamefont [1]{#1}%
\providecommand \citenamefont [1]{#1}%
\providecommand\href[0]{\@sanitize\@href}%
\providecommand\@href[1]{\endgroup\@@startlink{#1}\endgroup\@@href}%
\providecommand\@@href[1]{#1\@@endlink}%
\providecommand \@sanitize [0]{\begingroup\catcode`\&12\catcode`\#12\relax}%
\@ifxundefined \pdfoutput {\@firstoftwo}{%
 \@ifnum{\z@=\pdfoutput}{\@firstoftwo}{\@secondoftwo}%
}{%
 \providecommand\@@startlink[1]{\leavevmode\special{html:<a href="#1">}}%
 \providecommand\@@endlink[0]{\special{html:</a>}}%
}{%
 \providecommand\@@startlink[1]{%
  \leavevmode
  \pdfstartlink
   attr{/Border[0 0 1 ]/H/I/C[0 1 1]}%
   user{/Subtype/Link/A<</Type/Action/S/URI/URI(#1)>>}%
  \relax
 }%
 \providecommand\@@endlink[0]{\pdfendlink}%
}%
\providecommand \url  [0]{\begingroup\@sanitize \@url }%
\providecommand \@url [1]{\endgroup\@href {#1}{\urlprefix}}%
\providecommand \urlprefix [0]{URL }%
\providecommand \Eprint[0]{\href }%
\@ifxundefined \urlstyle {%
  \providecommand \doi [1]{doi:\discretionary{}{}{}#1}%
}{%
  \providecommand \doi [0]{doi:\discretionary{}{}{}\begingroup
  \urlstyle{rm}\Url }%
}%
\providecommand \doibase [0]{http://dx.doi.org/}%
\providecommand \Doi[1]{\href{\doibase#1}}%
\providecommand \bibAnnote [3]{%
  \BibitemShut{#1}%
  \begin{quotation}\noindent
    \textsc{Key:}\ #2\\\textsc{Annotation:}\ #3%
  \end{quotation}%
}%
\providecommand \bibAnnoteFile [2]{%
  \IfFileExists{#2}{\bibAnnote {#1} {#2} {\input{#2}}}{}%
}%
\providecommand \typeout [0]{\immediate \write \m@ne }%
\providecommand \selectlanguage [0]{\@gobble}%
\providecommand \bibinfo [0]{\@secondoftwo}%
\providecommand \bibfield [0]{\@secondoftwo}%
\providecommand \translation [1]{[#1]}%
\providecommand \BibitemOpen[0]{}%
\providecommand \bibitemStop [0]{}%
\providecommand \bibitemNoStop [0]{.\EOS\space}%
\providecommand \EOS [0]{\spacefactor3000\relax}%
\providecommand \BibitemShut [1]{\csname bibitem#1\endcsname}%
\bibitem{Hussey02}%
  \BibitemOpen
  \bibfield{author}{%
  \bibinfo {author} {\bibfnamefont{N.~E.}\ \bibnamefont{Hussey}},\ }%
  \bibfield{journal}{%
  \bibinfo {journal} {Adv. Phys.}\ }%
  \textbf{\bibinfo {volume} {51}},\ \bibinfo {pages} {1685} (\bibinfo {year}
  {2002})%
  \bibAnnoteFile{NoStop}{Hussey02}%
\bibitem{Fisher07}%
  \BibitemOpen
  \bibfield{author}{%
  \bibinfo {author} {\bibfnamefont{R.~A.}\ \bibnamefont{Fisher}}, \bibinfo
  {author} {\bibfnamefont{J.~E.}\ \bibnamefont{Gordon}},\ and\ \bibinfo
  {author} {\bibfnamefont{N.~E.}\ \bibnamefont{Phillips}},\ }%
  in\ \emph{\bibinfo {booktitle} {Handbook of High-Temperature
  Superconductivity: Theory and Experiment}},\ \bibinfo {editor} {edited by\
  \bibinfo {editor} {\bibfnamefont{J.~R.}\ \bibnamefont{Schrieffer}}\ and\
  \bibinfo {editor} {\bibfnamefont{J.~S.}\ \bibnamefont{Brooks}}}\ (\bibinfo
  {publisher} {Springer, New York},\ \bibinfo {year} {2007})\ Chap.~\bibinfo
  {chapter} {9}%
  \bibAnnoteFile{NoStop}{Fisher07}%
\bibitem{Volovik93}%
  \BibitemOpen
  \bibfield{author}{%
  \bibinfo {author} {\bibfnamefont{G.~E.}\ \bibnamefont{Volovik}},\ }%
  \bibfield{journal}{%
  \bibinfo {journal} {JETP Lett.}\ }%
  \textbf{\bibinfo {volume} {58}},\ \bibinfo {pages} {469} (\bibinfo {year}
  {1993})%
  \bibAnnoteFile{NoStop}{Volovik93}%
\bibitem{Loram01}%
  \BibitemOpen
  \bibfield{author}{%
  \bibinfo {author} {\bibfnamefont{J.~W.}\ \bibnamefont{Loram}}, \bibinfo
  {author} {\bibfnamefont{J.}~\bibnamefont{Luo}}, \bibinfo {author}
  {\bibfnamefont{J.~R.}\ \bibnamefont{Cooper}}, \bibinfo {author}
  {\bibfnamefont{W.~Y.}\ \bibnamefont{Liang}},\ and\ \bibinfo {author}
  {\bibfnamefont{J.~L.}\ \bibnamefont{Tallon}},\ }%
  \bibfield{journal}{%
  \bibinfo {journal} {J. Phys. Chem. Solids}\ }%
  \textbf{\bibinfo {volume} {62}},\ \bibinfo {pages} {59} (\bibinfo {year}
  {2001})%
  \bibAnnoteFile{NoStop}{Loram01}%
\bibitem{Wang01}%
  \BibitemOpen
  \bibfield{author}{%
  \bibinfo {author} {\bibfnamefont{Y.}~\bibnamefont{Wang}}, \bibinfo {author}
  {\bibfnamefont{B.}~\bibnamefont{Revaz}}, \bibinfo {author}
  {\bibfnamefont{A.}~\bibnamefont{Erb}},\ and\ \bibinfo {author}
  {\bibfnamefont{A.}~\bibnamefont{Junod}},\ }%
  \bibfield{journal}{%
  \bibinfo {journal} {Phys. Rev. B}\ }%
  \textbf{\bibinfo {volume} {63}},\ \bibinfo {pages} {094508} (\bibinfo {year}
  {2001})%
  \bibAnnoteFile{NoStop}{Wang01}%
\bibitem{Wen05}%
  \BibitemOpen
  \bibfield{author}{%
  \bibinfo {author} {\bibfnamefont{H.-H.}\ \bibnamefont{Wen}}, \bibinfo
  {author} {\bibfnamefont{L.}~\bibnamefont{Shan}}, \bibinfo {author}
  {\bibfnamefont{X.~G.}\ \bibnamefont{Wen}}, \bibinfo {author}
  {\bibfnamefont{Y.}~\bibnamefont{Wang}}, \bibinfo {author}
  {\bibfnamefont{H.}~\bibnamefont{Gao}}, \bibinfo {author}
  {\bibfnamefont{Z.-Y.}\ \bibnamefont{Liu}}, \bibinfo {author}
  {\bibfnamefont{F.}~\bibnamefont{Zhou}}, \bibinfo {author}
  {\bibfnamefont{J.~W.}\ \bibnamefont{Xiong}},\ and\ \bibinfo {author}
  {\bibfnamefont{W.~X.}\ \bibnamefont{Ti}},\ }%
  \bibfield{journal}{%
  \bibinfo {journal} {Phys. Rev. B}\ }%
  \textbf{\bibinfo {volume} {72}},\ \bibinfo {pages} {134507} (\bibinfo {year}
  {2005})%
  \bibAnnoteFile{NoStop}{Wen05}%
\bibitem{Wang07}%
  \BibitemOpen
  \bibfield{author}{%
  \bibinfo {author} {\bibfnamefont{Y.}~\bibnamefont{Wang}}, \bibinfo {author}
  {\bibfnamefont{J.}~\bibnamefont{Yan}}, \bibinfo {author}
  {\bibfnamefont{L.}~\bibnamefont{Shan}}, \bibinfo {author}
  {\bibfnamefont{H.-H.}\ \bibnamefont{Wen}}, \bibinfo {author}
  {\bibfnamefont{Y.}~\bibnamefont{Tanabe}}, \bibinfo {author}
  {\bibfnamefont{T.}~\bibnamefont{Adachi}},\ and\ \bibinfo {author}
  {\bibfnamefont{Y.}~\bibnamefont{Koike}},\ }%
  \bibfield{journal}{%
  \bibinfo {journal} {Phys. Rev. B}\ }%
  \textbf{\bibinfo {volume} {76}},\ \bibinfo {pages} {064512} (\bibinfo {year}
  {2007})%
  \bibAnnoteFile{NoStop}{Wang07}%
\bibitem{Harris97}%
  \BibitemOpen
  \bibfield{author}{%
  \bibinfo {author} {\bibfnamefont{J.~M.}\ \bibnamefont{Harris}}, \bibinfo
  {author} {\bibfnamefont{P.~J.}\ \bibnamefont{White}}, \bibinfo {author}
  {\bibfnamefont{Z.-X.}\ \bibnamefont{Shen}}, \bibinfo {author}
  {\bibfnamefont{H.}~\bibnamefont{Ikeda}}, \bibinfo {author}
  {\bibfnamefont{R.}~\bibnamefont{Yoshizaki}}, \bibinfo {author}
  {\bibfnamefont{H.}~\bibnamefont{Eisaki}}, \bibinfo {author}
  {\bibfnamefont{S.}~\bibnamefont{Uchida}}, \bibinfo {author}
  {\bibfnamefont{W.~D.}\ \bibnamefont{Si}}, \bibinfo {author}
  {\bibfnamefont{J.~W.}\ \bibnamefont{Xiong}}, \bibinfo {author}
  {\bibfnamefont{Z.-X.}\ \bibnamefont{Zhao}},\ and\ \bibinfo {author}
  {\bibfnamefont{D.~S.}\ \bibnamefont{Dessau}},\ }%
  \bibfield{journal}{%
  \bibinfo {journal} {Phys. Rev. Lett.}\ }%
  \textbf{\bibinfo {volume} {79}},\ \bibinfo {pages} {143} (\bibinfo {year}
  {1997})%
  \bibAnnoteFile{NoStop}{Harris97}%
\bibitem{Kondo07}%
  \BibitemOpen
  \bibfield{author}{%
  \bibinfo {author} {\bibfnamefont{T.}~\bibnamefont{Kondo}}, \bibinfo {author}
  {\bibfnamefont{T.}~\bibnamefont{Takeuchi}}, \bibinfo {author}
  {\bibfnamefont{A.}~\bibnamefont{Kaminski}}, \bibinfo {author}
  {\bibfnamefont{S.}~\bibnamefont{Tsuda}},\ and\ \bibinfo {author}
  {\bibfnamefont{S.}~\bibnamefont{Shin}},\ }%
  \bibfield{journal}{%
  \bibinfo {journal} {Phys. Rev. Lett.}\ }%
  \textbf{\bibinfo {volume} {98}},\ \bibinfo {pages} {267004} (\bibinfo {year}
  {2007})%
  \bibAnnoteFile{NoStop}{Kondo07}%
\bibitem{Wei08}%
  \BibitemOpen
  \bibfield{author}{%
  \bibinfo {author} {\bibfnamefont{J.}~\bibnamefont{Wei}}, \bibinfo {author}
  {\bibfnamefont{Y.}~\bibnamefont{Zhang}}, \bibinfo {author}
  {\bibfnamefont{H.~W.}\ \bibnamefont{Ou}}, \bibinfo {author}
  {\bibfnamefont{B.~P.}\ \bibnamefont{Xie}}, \bibinfo {author}
  {\bibfnamefont{D.~W.}\ \bibnamefont{Shen}}, \bibinfo {author}
  {\bibfnamefont{J.~F.}\ \bibnamefont{Zhao}}, \bibinfo {author}
  {\bibfnamefont{L.~X.}\ \bibnamefont{Yang}}, \bibinfo {author}
  {\bibfnamefont{M.}~\bibnamefont{Arita}}, \bibinfo {author}
  {\bibfnamefont{K.}~\bibnamefont{Shimada}}, \bibinfo {author}
  {\bibfnamefont{H.}~\bibnamefont{Namatame}}, \bibinfo {author}
  {\bibfnamefont{M.}~\bibnamefont{Taniguchi}}, \bibinfo {author}
  {\bibfnamefont{Y.}~\bibnamefont{Yoshida}}, \bibinfo {author}
  {\bibfnamefont{H.}~\bibnamefont{Eisaki}},\ and\ \bibinfo {author}
  {\bibfnamefont{D.~L.}\ \bibnamefont{Feng}},\ }%
  \bibfield{journal}{%
  \bibinfo {journal} {Phys. Rev. Lett.}\ }%
  \textbf{\bibinfo {volume} {101}},\ \bibinfo {pages} {097005} (\bibinfo {year}
  {2008})%
  \bibAnnoteFile{NoStop}{Wei08}%
\bibitem{Ma08}%
  \BibitemOpen
  \bibfield{author}{%
  \bibinfo {author} {\bibfnamefont{J.-H.}\ \bibnamefont{Ma}}, \bibinfo {author}
  {\bibfnamefont{Z.-H.}\ \bibnamefont{Pan}}, \bibinfo {author}
  {\bibfnamefont{F.~C.}\ \bibnamefont{Niestemski}}, \bibinfo {author}
  {\bibfnamefont{M.}~\bibnamefont{Neupane}}, \bibinfo {author}
  {\bibfnamefont{Y.-M.}\ \bibnamefont{Xu}}, \bibinfo {author}
  {\bibfnamefont{P.}~\bibnamefont{Richard}}, \bibinfo {author}
  {\bibfnamefont{K.}~\bibnamefont{Nakayama}}, \bibinfo {author}
  {\bibfnamefont{T.}~\bibnamefont{Sato}}, \bibinfo {author}
  {\bibfnamefont{T.}~\bibnamefont{Takahashi}}, \bibinfo {author}
  {\bibfnamefont{H.-Q.}\ \bibnamefont{Luo}}, \bibinfo {author}
  {\bibfnamefont{L.}~\bibnamefont{Fang}}, \bibinfo {author}
  {\bibfnamefont{H.-H.}\ \bibnamefont{Wen}}, \bibinfo {author}
  {\bibfnamefont{Z.}~\bibnamefont{Wang}}, \bibinfo {author}
  {\bibfnamefont{H.}~\bibnamefont{Ding}},\ and\ \bibinfo {author}
  {\bibfnamefont{V.}~\bibnamefont{Madhavan}},\ }%
  \bibfield{journal}{%
  \bibinfo {journal} {Phys. Rev. Lett.}\ }%
  \textbf{\bibinfo {volume} {101}},\ \bibinfo {pages} {207002} (\bibinfo {year}
  {2008})%
  \bibAnnoteFile{NoStop}{Ma08}%
\bibitem{Meng09}%
  \BibitemOpen
  \bibfield{author}{%
  \bibinfo {author} {\bibfnamefont{J.~Q.}\ \bibnamefont{Meng}}, \bibinfo
  {author} {\bibfnamefont{W.~T.}\ \bibnamefont{Zhang}}, \bibinfo {author}
  {\bibfnamefont{G.~D.}\ \bibnamefont{Liu}}, \bibinfo {author}
  {\bibfnamefont{L.}~\bibnamefont{Zhao}}, \bibinfo {author}
  {\bibfnamefont{H.~Y.}\ \bibnamefont{Liu}}, \bibinfo {author}
  {\bibfnamefont{X.~W.}\ \bibnamefont{Jia}}, \bibinfo {author}
  {\bibfnamefont{W.}~\bibnamefont{Lu}}, \bibinfo {author}
  {\bibfnamefont{X.~L.}\ \bibnamefont{Dong}}, \bibinfo {author}
  {\bibfnamefont{G.~L.}\ \bibnamefont{Wang}}, \bibinfo {author}
  {\bibfnamefont{H.~B.}\ \bibnamefont{Zhang}}, \bibinfo {author}
  {\bibfnamefont{Y.}~\bibnamefont{Zhou}}, \bibinfo {author}
  {\bibfnamefont{Y.}~\bibnamefont{Zhu}}, \bibinfo {author}
  {\bibfnamefont{X.~Y.}\ \bibnamefont{Wang}}, \bibinfo {author}
  {\bibfnamefont{Z.-X.}\ \bibnamefont{Zhao}}, \bibinfo {author}
  {\bibfnamefont{Z.~Y.}\ \bibnamefont{Xu}}, \bibinfo {author}
  {\bibfnamefont{C.~T.}\ \bibnamefont{Chen}},\ and\ \bibinfo {author}
  {\bibfnamefont{X.~J.}\ \bibnamefont{Zhou}},\ }%
  \bibfield{journal}{%
  \bibinfo {journal} {Phys. Rev. B}\ }%
  \textbf{\bibinfo {volume} {79}},\ \bibinfo {pages} {024514} (\bibinfo {year}
  {2009})%
  \bibAnnoteFile{NoStop}{Meng09}%
\bibitem{Liang04}%
  \BibitemOpen
  \bibfield{author}{%
  \bibinfo {author} {\bibfnamefont{B.}~\bibnamefont{Liang}}\ and\ \bibinfo
  {author} {\bibfnamefont{C.~T.}\ \bibnamefont{Lin}},\ }%
  \bibfield{journal}{%
  \bibinfo {journal} {J. Cryst. Growth}\ }%
  \textbf{\bibinfo {volume} {267}},\ \bibinfo {pages} {510} (\bibinfo {year}
  {2004})%
  \bibAnnoteFile{NoStop}{Liang04}%
\bibitem{Note1}%
  \BibitemOpen
  \bibinfo {note} {The actual La content of the sample is inferred to be 0.36,
  based on the wavelength dispersive spectroscopy (WDS) measurements on
  crystals grown under the same condition (Ref. \onlinecite{Liang04}). With
  this, the sample's hole doping level $p$ is estimated to be $\sim$0.165
  according to the empirical relation between $x$ and $p$ proposed in the
  literature (Refs. \onlinecite{Ando00,Luo08}). This slight deviation of the
  $p$ from the optimal doping point (0.16) may explain the $T_c$ of the sample
  lower than $T_c^{\mathrm{max}}\simeq32$ K.}%
  \bibAnnoteFile{Stop}{Note1}%
\bibitem{Wen04}%
  \BibitemOpen
  \bibfield{author}{%
  \bibinfo {author} {\bibfnamefont{H.-H.}\ \bibnamefont{Wen}}, \bibinfo
  {author} {\bibfnamefont{Z.-Y.}\ \bibnamefont{Liu}}, \bibinfo {author}
  {\bibfnamefont{F.}~\bibnamefont{Zhou}}, \bibinfo {author}
  {\bibfnamefont{J.~W.}\ \bibnamefont{Xiong}}, \bibinfo {author}
  {\bibfnamefont{W.~X.}\ \bibnamefont{Ti}}, \bibinfo {author}
  {\bibfnamefont{T.}~\bibnamefont{Xiang}}, \bibinfo {author}
  {\bibfnamefont{S.}~\bibnamefont{Komiya}}, \bibinfo {author}
  {\bibfnamefont{X.~F.}\ \bibnamefont{Sun}},\ and\ \bibinfo {author}
  {\bibfnamefont{Y.}~\bibnamefont{Ando}},\ }%
  \bibfield{journal}{%
  \bibinfo {journal} {Phys. Rev. B}\ }%
  \textbf{\bibinfo {volume} {70}},\ \bibinfo {pages} {214505} (\bibinfo {year}
  {2004})%
  \bibAnnoteFile{NoStop}{Wen04}%
\bibitem{Moler97}%
  \BibitemOpen
  \bibfield{author}{%
  \bibinfo {author} {\bibfnamefont{K.~A.}\ \bibnamefont{Moler}}, \bibinfo
  {author} {\bibfnamefont{D.~L.}\ \bibnamefont{Sisson}}, \bibinfo {author}
  {\bibfnamefont{J.~S.}\ \bibnamefont{Urbach}}, \bibinfo {author}
  {\bibfnamefont{M.~R.}\ \bibnamefont{Beasley}}, \bibinfo {author}
  {\bibfnamefont{A.}~\bibnamefont{Kapitulnik}}, \bibinfo {author}
  {\bibfnamefont{D.~J.}\ \bibnamefont{Baar}}, \bibinfo {author}
  {\bibfnamefont{R.}~\bibnamefont{Liang}},\ and\ \bibinfo {author}
  {\bibfnamefont{W.~N.}\ \bibnamefont{Hardy}},\ }%
  \bibfield{journal}{%
  \bibinfo {journal} {Phys. Rev. B}\ }%
  \textbf{\bibinfo {volume} {55}},\ \bibinfo {pages} {3954} (\bibinfo {year}
  {1997})%
  \bibAnnoteFile{NoStop}{Moler97}%
\bibitem{Fisher00}%
  \BibitemOpen
  \bibfield{author}{%
  \bibinfo {author} {\bibfnamefont{R.~A.}\ \bibnamefont{Fisher}}, \bibinfo
  {author} {\bibfnamefont{N.~E.}\ \bibnamefont{Phillips}}, \bibinfo {author}
  {\bibfnamefont{A.}~\bibnamefont{Schilling}}, \bibinfo {author}
  {\bibfnamefont{B.}~\bibnamefont{Buffeteau}}, \bibinfo {author}
  {\bibfnamefont{R.}~\bibnamefont{Calemczuk}}, \bibinfo {author}
  {\bibfnamefont{T.~E.}\ \bibnamefont{Hargreaves}}, \bibinfo {author}
  {\bibfnamefont{C.}~\bibnamefont{Marcenat}}, \bibinfo {author}
  {\bibfnamefont{K.~W.}\ \bibnamefont{Dennis}}, \bibinfo {author}
  {\bibfnamefont{R.~W.}\ \bibnamefont{McCallum}},\ and\ \bibinfo {author}
  {\bibfnamefont{A.~S.}\ \bibnamefont{O'Connor}},\ }%
  \bibfield{journal}{%
  \bibinfo {journal} {Phys. Rev. B}\ }%
  \textbf{\bibinfo {volume} {61}},\ \bibinfo {pages} {1473} (\bibinfo {year}
  {2000})%
  \bibAnnoteFile{NoStop}{Fisher00}%
\bibitem{Chakraborty89}%
  \BibitemOpen
  \bibfield{author}{%
  \bibinfo {author} {\bibfnamefont{A.}~\bibnamefont{Chakraborty}}, \bibinfo
  {author} {\bibfnamefont{A.~J.}\ \bibnamefont{Epstein}}, \bibinfo {author}
  {\bibfnamefont{D.~L.}\ \bibnamefont{Cox}}, \bibinfo {author}
  {\bibfnamefont{E.~M.}\ \bibnamefont{McCarron}},\ and\ \bibinfo {author}
  {\bibfnamefont{W.~E.}\ \bibnamefont{Farneth}},\ }%
  \bibfield{journal}{%
  \bibinfo {journal} {Phys. Rev. B}\ }%
  \textbf{\bibinfo {volume} {39}},\ \bibinfo {pages} {12267} (\bibinfo {year}
  {1989})%
  \bibAnnoteFile{NoStop}{Chakraborty89}%
\bibitem{Urbach89}%
  \BibitemOpen
  \bibfield{author}{%
  \bibinfo {author} {\bibfnamefont{J.~S.}\ \bibnamefont{Urbach}}, \bibinfo
  {author} {\bibfnamefont{D.~B.}\ \bibnamefont{Mitzi}}, \bibinfo {author}
  {\bibfnamefont{A.}~\bibnamefont{Kapitulnik}}, \bibinfo {author}
  {\bibfnamefont{J.~Y.~T.}\ \bibnamefont{Wei}},\ and\ \bibinfo {author}
  {\bibfnamefont{D.~E.}\ \bibnamefont{Morris}},\ }%
  \bibfield{journal}{%
  \bibinfo {journal} {Phys. Rev. B}\ }%
  \textbf{\bibinfo {volume} {39}},\ \bibinfo {pages} {12391} (\bibinfo {year}
  {1989})%
  \bibAnnoteFile{NoStop}{Urbach89}%
\bibitem{Note2}%
  \BibitemOpen
  \bibinfo {note} {As shown in Fig. \ref{figure2}, there is no discernible
  Schottky-like anomaly in all LTSH data sets at different $H$, which has
  helped simplify the data analysis.}%
  \bibAnnoteFile{Stop}{Note2}%
\bibitem{Chiao00}%
  \BibitemOpen
  \bibfield{author}{%
  \bibinfo {author} {\bibfnamefont{M.}~\bibnamefont{Chiao}}, \bibinfo {author}
  {\bibfnamefont{R.~W.}\ \bibnamefont{Hill}}, \bibinfo {author}
  {\bibfnamefont{C.}~\bibnamefont{Lupien}}, \bibinfo {author}
  {\bibfnamefont{L.}~\bibnamefont{Taillefer}}, \bibinfo {author}
  {\bibfnamefont{P.}~\bibnamefont{Lambert}}, \bibinfo {author}
  {\bibfnamefont{R.}~\bibnamefont{Gagnon}},\ and\ \bibinfo {author}
  {\bibfnamefont{P.}~\bibnamefont{Fournier}},\ }%
  \bibfield{journal}{%
  \bibinfo {journal} {Phys. Rev. B}\ }%
  \textbf{\bibinfo {volume} {62}},\ \bibinfo {pages} {3554} (\bibinfo {year}
  {2000})%
  \bibAnnoteFile{NoStop}{Chiao00}%
\bibitem{Vekhter01}%
  \BibitemOpen
  \bibfield{author}{%
  \bibinfo {author} {\bibfnamefont{I.}~\bibnamefont{Vekhter}}, \bibinfo
  {author} {\bibfnamefont{P.~J.}\ \bibnamefont{Hirschfeld}},\ and\ \bibinfo
  {author} {\bibfnamefont{E.~J.}\ \bibnamefont{Nicol}},\ }%
  \bibfield{journal}{%
  \bibinfo {journal} {Phys. Rev. B}\ }%
  \textbf{\bibinfo {volume} {64}},\ \bibinfo {pages} {064513} (\bibinfo {year}
  {2001})%
  \bibAnnoteFile{NoStop}{Vekhter01}%
\bibitem{Hashimoto08}%
  \BibitemOpen
  \bibfield{author}{%
  \bibinfo {author} {\bibfnamefont{M.}~\bibnamefont{Hashimoto}}, \bibinfo
  {author} {\bibfnamefont{T.}~\bibnamefont{Yoshida}}, \bibinfo {author}
  {\bibfnamefont{H.}~\bibnamefont{Yagi}}, \bibinfo {author}
  {\bibfnamefont{M.}~\bibnamefont{Takizawa}}, \bibinfo {author}
  {\bibfnamefont{A.}~\bibnamefont{Fujimori}}, \bibinfo {author}
  {\bibfnamefont{M.}~\bibnamefont{Kubota}}, \bibinfo {author}
  {\bibfnamefont{K.}~\bibnamefont{Ono}}, \bibinfo {author}
  {\bibfnamefont{K.}~\bibnamefont{Tanaka}}, \bibinfo {author}
  {\bibfnamefont{D.~H.}\ \bibnamefont{Lu}}, \bibinfo {author}
  {\bibfnamefont{Z.-X.}\ \bibnamefont{Shen}}, \bibinfo {author}
  {\bibfnamefont{S.}~\bibnamefont{Ono}},\ and\ \bibinfo {author}
  {\bibfnamefont{Y.}~\bibnamefont{Ando}},\ }%
  \bibfield{journal}{%
  \bibinfo {journal} {Phys. Rev. B}\ }%
  \textbf{\bibinfo {volume} {77}},\ \bibinfo {pages} {094516} (\bibinfo {year}
  {2008})%
  \bibAnnoteFile{NoStop}{Hashimoto08}%
\bibitem{Won94}%
  \BibitemOpen
  \bibfield{author}{%
  \bibinfo {author} {\bibfnamefont{H.}~\bibnamefont{Won}}\ and\ \bibinfo
  {author} {\bibfnamefont{K.}~\bibnamefont{Maki}},\ }%
  \bibfield{journal}{%
  \bibinfo {journal} {Phys. Rev. B}\ }%
  \textbf{\bibinfo {volume} {49}},\ \bibinfo {pages} {1397} (\bibinfo {year}
  {1994})%
  \bibAnnoteFile{NoStop}{Won94}%
\bibitem{Nakayama07}%
  \BibitemOpen
  \bibfield{author}{%
  \bibinfo {author} {\bibfnamefont{K.}~\bibnamefont{Nakayama}}, \bibinfo
  {author} {\bibfnamefont{T.}~\bibnamefont{Sato}}, \bibinfo {author}
  {\bibfnamefont{K.}~\bibnamefont{Terashima}}, \bibinfo {author}
  {\bibfnamefont{H.}~\bibnamefont{Matsui}}, \bibinfo {author}
  {\bibfnamefont{T.}~\bibnamefont{Takahashi}}, \bibinfo {author}
  {\bibfnamefont{M.}~\bibnamefont{Kubota}}, \bibinfo {author}
  {\bibfnamefont{K.}~\bibnamefont{Ono}}, \bibinfo {author}
  {\bibfnamefont{T.}~\bibnamefont{Nishizaki}}, \bibinfo {author}
  {\bibfnamefont{Y.}~\bibnamefont{Takahashi}},\ and\ \bibinfo {author}
  {\bibfnamefont{N.}~\bibnamefont{Kobayashi}},\ }%
  \bibfield{journal}{%
  \bibinfo {journal} {Phys. Rev. B}\ }%
  \textbf{\bibinfo {volume} {75}},\ \bibinfo {pages} {014513} (\bibinfo {year}
  {2007})%
  \bibAnnoteFile{NoStop}{Nakayama07}%
\bibitem{Ding96}%
  \BibitemOpen
  \bibfield{author}{%
  \bibinfo {author} {\bibfnamefont{H.}~\bibnamefont{Ding}}, \bibinfo {author}
  {\bibfnamefont{M.~R.}\ \bibnamefont{Norman}}, \bibinfo {author}
  {\bibfnamefont{J.~C.}\ \bibnamefont{Campuzano}}, \bibinfo {author}
  {\bibfnamefont{M.}~\bibnamefont{Randeria}}, \bibinfo {author}
  {\bibfnamefont{A.~F.}\ \bibnamefont{Bellman}}, \bibinfo {author}
  {\bibfnamefont{T.}~\bibnamefont{Yokoya}}, \bibinfo {author}
  {\bibfnamefont{T.}~\bibnamefont{Takahashi}}, \bibinfo {author}
  {\bibfnamefont{T.}~\bibnamefont{Mochiku}},\ and\ \bibinfo {author}
  {\bibfnamefont{K.}~\bibnamefont{Kadowaki}},\ }%
  \bibfield{journal}{%
  \bibinfo {journal} {Phys. Rev. B}\ }%
  \textbf{\bibinfo {volume} {54}},\ \bibinfo {pages} {R9678} (\bibinfo {year}
  {1996})%
  \bibAnnoteFile{NoStop}{Ding96}%
\bibitem{Ando00}%
  \BibitemOpen
  \bibfield{author}{%
  \bibinfo {author} {\bibfnamefont{Y.}~\bibnamefont{Ando}}, \bibinfo {author}
  {\bibfnamefont{Y.}~\bibnamefont{Hanaki}}, \bibinfo {author}
  {\bibfnamefont{S.}~\bibnamefont{Ono}}, \bibinfo {author}
  {\bibfnamefont{T.}~\bibnamefont{Murayama}}, \bibinfo {author}
  {\bibfnamefont{K.}~\bibnamefont{Segawa}}, \bibinfo {author}
  {\bibfnamefont{N.}~\bibnamefont{Miyamoto}},\ and\ \bibinfo {author}
  {\bibfnamefont{S.}~\bibnamefont{Komiya}},\ }%
  \bibfield{journal}{%
  \bibinfo {journal} {Phys. Rev. B}\ }%
  \textbf{\bibinfo {volume} {61}},\ \bibinfo {pages} {R14956} (\bibinfo {year}
  {2000})%
  \bibAnnoteFile{NoStop}{Ando00}%
\bibitem{Luo08}%
  \BibitemOpen
  \bibfield{author}{%
  \bibinfo {author} {\bibfnamefont{H.-Q.}\ \bibnamefont{Luo}}, \bibinfo
  {author} {\bibfnamefont{P.}~\bibnamefont{Cheng}}, \bibinfo {author}
  {\bibfnamefont{L.}~\bibnamefont{Fang}},\ and\ \bibinfo {author}
  {\bibfnamefont{H.-H.}\ \bibnamefont{Wen}},\ }%
  \bibfield{journal}{%
  \bibinfo {journal} {Supercond. Sci. Technol.}\ }%
  \textbf{\bibinfo {volume} {21}},\ \bibinfo {pages} {125024} (\bibinfo {year}
  {2008})%
  \bibAnnoteFile{NoStop}{Luo08}%
\bibitem{Shi08}%
  \BibitemOpen
  \bibfield{author}{%
  \bibinfo {author} {\bibfnamefont{M.}~\bibnamefont{Shi}}, \bibinfo {author}
  {\bibfnamefont{J.}~\bibnamefont{Chang}}, \bibinfo {author}
  {\bibfnamefont{S.}~\bibnamefont{Pailh\'es}}, \bibinfo {author}
  {\bibfnamefont{M.~R.}\ \bibnamefont{Norman}}, \bibinfo {author}
  {\bibfnamefont{J.~C.}\ \bibnamefont{Campuzano}}, \bibinfo {author}
  {\bibfnamefont{M.}~\bibnamefont{M{\aa}nsson}}, \bibinfo {author}
  {\bibfnamefont{T.}~\bibnamefont{Claesson}}, \bibinfo {author}
  {\bibfnamefont{O.}~\bibnamefont{Tjernberg}}, \bibinfo {author}
  {\bibfnamefont{A.}~\bibnamefont{Bendounan}}, \bibinfo {author}
  {\bibfnamefont{L.}~\bibnamefont{Patthey}}, \bibinfo {author}
  {\bibfnamefont{N.}~\bibnamefont{Momono}}, \bibinfo {author}
  {\bibfnamefont{M.}~\bibnamefont{Oda}}, \bibinfo {author}
  {\bibfnamefont{M.}~\bibnamefont{Ido}}, \bibinfo {author}
  {\bibfnamefont{C.}~\bibnamefont{Mudry}},\ and\ \bibinfo {author}
  {\bibfnamefont{J.}~\bibnamefont{Mesot}},\ }%
  \bibfield{journal}{%
  \bibinfo {journal} {Phys. Rev. Lett.}\ }%
  \textbf{\bibinfo {volume} {101}},\ \bibinfo {pages} {047002} (\bibinfo {year}
  {2008})%
  \bibAnnoteFile{NoStop}{Shi08}%
\bibitem{Nakano98}%
  \BibitemOpen
  \bibfield{author}{%
  \bibinfo {author} {\bibfnamefont{T.}~\bibnamefont{Nakano}}, \bibinfo {author}
  {\bibfnamefont{N.}~\bibnamefont{Momono}}, \bibinfo {author}
  {\bibfnamefont{M.}~\bibnamefont{Oda}},\ and\ \bibinfo {author}
  {\bibfnamefont{M.}~\bibnamefont{Ido}},\ }%
  \bibfield{journal}{%
  \bibinfo {journal} {J. Phys. Soc. Jpn.}\ }%
  \textbf{\bibinfo {volume} {67}},\ \bibinfo {pages} {2622} (\bibinfo {year}
  {1998})%
  \bibAnnoteFile{NoStop}{Nakano98}%
\bibitem{Millis06}%
  \BibitemOpen
  \bibfield{author}{%
  \bibinfo {author} {\bibfnamefont{A.~J.}\ \bibnamefont{Millis}},\ }%
  \bibfield{journal}{%
  \bibinfo {journal} {Science}\ }%
  \textbf{\bibinfo {volume} {314}},\ \bibinfo {pages} {1888} (\bibinfo {year}
  {2006})%
  \bibAnnoteFile{NoStop}{Millis06}%
\end{thebibliography}%

\end{document}